\documentclass[letterpaper]{emulateapj}
\usepackage{amsmath}
\usepackage{graphics}
\usepackage{amsmath,cases}
\usepackage{color}
\newcommand{\msun}{{M}_{\odot}}
\newcommand{\rsun}{{R}_{\odot}}
\newcommand{\zsun}{{Z}_{\odot}}

\shorttitle{GRB RATE in Merger Model}
\shortauthors{KINUGAWA \& ASANO.}

\begin{document}

	
	\title{LONG Gamma-ray burst RATE in the BINARY MERGER progenitor MODEL}
	

	\author{Tomoya Kinugawa\altaffilmark{1} and Katsuaki Asano\altaffilmark{1}}
	%
	%

\altaffiltext{1}{Institute for Cosmic Ray Research, The University of Tokyo, 5-1-5 Kashiwa-no-ha, Kashiwa City, Chiba, 277-8582, Japan}

\begin{abstract}
	The long gamma-ray bursts (GRBs) may arise from the core collapse of massive stars.
	However, the  long GRB rate does not follow the star formation rate (SFR) at  high redshifts.
	In this {\it Letter}, we focus on the binary merger model and consider the high spin helium stars after the merger as the progenitor of long GRBs.
With this scenario, we estimate the GRB
rate by the population synthesis method
with the metallicity evolution.	
It is easier for low metallicity binaries to become long GRB progenitors than those of solar metallicity due to the weak wind mass loss and the difference in the stellar evolution.
In our results, the long GRB rate roughly agrees with the observed rate, and shows a similar behavior to the observed redshift evolution.
\end{abstract}
\section{introduction}
{The long gamma-ray burst (GRB) rate has been estimated with the {\it Swift} sample \cite[e.g.][]{WP10,Lien2014,Lien2015}.
The long GRBs may arise from the core collapse of  massive stars.
However, the derived redshift evolution of the GRB rate does not follow the star formation rate (SFR) especially at high redshift.
If the long GRB rate traces the SFR, the luminosity function (LF) should strongly evolve with the redshift to reconcile with the observations \citep{Lien2014,Pescalli2016}. 
Alternatively, the fraction of massive stars fated to yield a GRB may increase with redshift (e.g. Robertson \& Ellis 2012). The metallicity evolution is the most promising effect that affects the GRB rate.}

The black hole (BH) and accretion disk system in the core collapse of massive stars has been considered to be the central engine of long GRBs \citep{Woosley1993, McFadyen1999}.
In order to make the accretion disk around the BH, the progenitor star just before the collapse must have a high angular momentum \citep{Woosley2006,Yoon2006}.
However, massive stars generally lose a lot of angular momentum via the stellar wind mass loss.
{In order to overcome the angular momentum problem, several progenitor models have been proposed, such as the low metallicity star \citep{Hirschi2005, Yoon2005, Woosley2006, Yoon2006} or the binary merger progenitor models \citep{Fryer2005,Tout2011}, including models with specific evolution paths (e.g. \citet{Detmers2008}; tidal spin-up, \citet{P2010}; the explosive common-envelope ejection).}
In the low metallicity stars, the stellar wind mass loss is so weak that the stars hardly lose the spin  angular momentum.
 {However, in this case, the hydrogen envelope remains unless the star evolves as the chemically homogeneous evolution \citep{Maeder1987} with an extremely high spin, so the GRB jet may not penetrate the hydrogen envelope.}   
On the other hand, in the binary merger progenitor model, a Wolf-Rayet star and a giant star system or two giant stars merge during the common-envelope (CE) phase \citep{Webbink1984}.
 {When a star in a close binary system becomes a red giant, the envelope of the giant fills the Roche lobe. In this case, the mass transfer tends to be dynamically unstable; the subsequent evolutions of the radius of the giant and orbital motion further enhance the mass transfer rate.
Then, the companion star, which should be a post main-sequence (post-MS) star in the binary progenitor model, plunges into the envelope of the primary giant.}
In the CE phase, the companion star spirals into the core of the giant owing to the orbital energy loss by the friction.
After the CE phase, the envelope of the giant is evaporated and the binary separation becomes so close that they merge.
After the merger, a rapidly rotating naked helium star remains,  because the star obtains a lot of angular momentum from the orbital angular momentum at the merger.
Such a compact and highly rotating star is ideal to induce a GRB.

{In this paper, we consider the binary merger progenitor model \citep{Fryer2005} and estimate the GRB rate following the binary population synthesis method with the metallicity evolution.
The stellar wind mass loss reduces not only the spin angular momentum but also the orbital binding energy. 
If the metallicity of the binary system is high, the binary tends to become a wide system, so that the binary interaction may not work efficiently.
We can expect that a lower metallicity enhances the GRB production rate in this scenario.}

\section{method}

\begin{table*}[!ht]
	\caption{The initial distribution functions in This paper.
	}
	\label{IDF}
	\begin{center}
		\begin{tabular}{cccc}
			\hline
			IMF	& Initial Mass Ratio Function & 	Initial Period Function  &Initial Eccentricity function\\
			\hline
			Salpeter   &  $q^{-0.1}$&  ($\log P)^{-0.55}$ &$e^{-0.5}$\\
			$5~\msun<M_1<100~\msun$&$0.1~\msun/M_1<M_2/M_1<1$&$P_{\rm min}$*$<P<P_{\rm max}$&$0<e<1$\\
			\hline
		\end{tabular}\\
		{* We choose $P_{min}$ and $P_{\rm max}$ as the minimum period when the binary does not fulfill the Roche lobe \citep{Kinugawa2014} and the period when the separation is equal to $10^6~\rsun$.}
	\end{center}
\end{table*}

In order to calculate the long GRB rate based on the binary merger model, we use the population synthesis method.
{We calculate the binary evolutions for given initial binary parameters (the primary mass, the mass ratio, the orbital separation, and the eccentricity); we follow the evolutions of the stellar radius, the core mass, and the stellar wind mass loss and check whether the binary interaction occurs or not.}
With the Monte Carlo method under the initial distribution functions of the binary parameters, we estimate the fraction of the GRB progenitor systems for each metallicity. 
For the binary population synthesis code, we revise the BSE code \citep{Hurley2002}.
The wind mass-loss rate is the same as that in \cite{Kinugawa2017}.
In this mass-loss rate, the formulae are almost the same as those in the BSE code \citep{Hurley2000,Hurley2002}.
	However, we rewrited the metallicity dependence of the Wolf-Rayet star's mass-loss rate \citep{Vink2005} and the mass-loss rate for {LBV stars of Belczynski et al. (2010), in which the rate is chosen to reproduce the typical BH mass in our galaxy based on Humphreys and Davidson (1994). }

The CE phase is treated with the conventional method as follows.
The criterion of the dynamical instability to induce the CE phase is also the same as that in our previous papers \citep{Kinugawa2014,Kinugawa2016}.
The separation after the CE phase $a_{\rm f}$ is calculated using the energy balance prescription  \citep{Webbink1984}
\begin{equation}
\alpha\left(\frac{GM_{\rm{c,1}}M_2}{2a_{\rm{f}}}-\frac{GM_1M_2}{2a_{\rm{i}}}\right)=\frac{GM_{\rm{1}}M_{\rm{env,1}}}{\lambda R_1},
\label{eq:ce1}
\end{equation} 
where $a_{\rm i}$ is the binary separation just before the CE phase, and
$R_1$, $M_1$, $M_{\rm c,1}$, $M_{\rm env,1}$, and $M_2$ are the radius, the mass, the core mass and the envelope mass of the giant, and the mass of the companion star, respectively.
{When the companion star is also a giant, Equation (1) changes into
\begin{equation}
\alpha\left(\frac{GM_{\rm{c,1}}M_{c,2}}{2a_{\rm{f}}}-\frac{GM_1M_2}{2a_{\rm{i}}}\right)=\frac{GM_{\rm{1}}M_{\rm{env,1}}}{\lambda R_1}+\frac{GM_{\rm{2}}M_{\rm{env,2}}}{\lambda R_2},
\label{eq:ce2}
\end{equation}
where  $R_2$, $M_{\rm c,2}$, and $M_{\rm env,2}$ are  the radius, the mass, the core mass and the envelope mass of the companion star, respectively {(see also
\cite{Dewi2006}).} }
Although, the CE parameters $\alpha$ and $\lambda$ are not well understood \citep{Ivanova2013}, here we adopt $\alpha\lambda=1$ and {$0.1$.
A larger $\alpha \lambda$ leads to a wider separation after the CE phase, and vice versa.}
{Generally, the energies of the stellar wind and the radiation from WR stars are much larger than the binding energy of the envelope \citep[e.g.][]{Maeder2009}. Thus, even if the envelope ejected in the CE phase is fallen back,  it will possibly be evaporated by the stellar wind and the radiation.
In this paper, we optimistically assume that all the envelope evaporates during the CE phase.}

After the CE phase, if the separation is less than the sum of the core radius of the giant and the radius of the companion {(or the core of the companion if the companion is also a giant)}, the binary merges.
{On the other hand, when the post-MS star did not reach the Hayashi track or ignite helium burning, such a star, so-called  the Hertzsprung gap (HG) star, {may not} have a clear core-envelope structure. In this case, we assume the binary always merges in the CE phase \citep{Taam2000, Ivanova2004, Belczynski2008}.}

{The required naked helium star is produced from the mergers in the CE phase, which occurs from the mass transfer between a WR star and a post-MS star, or two post-MS stars. The possible post-MS stars are an HG star, a red giant (RG), a helium core burning (HeB) phase star, or an asymptotic giant branch (AGB) star which is in the second expansion phase after the end of HeB phase. 
When the helium cores do not ignite yet (e.g, HG stars, RGs), the mass transfer at this stage is called case B \citep{K1967}.
In this case, the helium burning of the merged star starts at the merger.
The zero-age naked helium star evolution of the Hurley code \citep{Hurley2000} is adopted to follow the evolution of the merged stars.
In case B, the lifetime of the merged helium star is long ($\sim 10^6$ years).
On the other hand, when the helium core of the primary star already ignites (e.g. HeB stars, AGB stars), the mass transfer at this stage is called case C \citep{L1970}.
In this case, the evolution of the merged helium star with the CO core, where the central helium burning may proceed, is calculated with the same method in \cite{Hurley2000, Hurley2002}.
For the merged stars produced in case C, the lifetime tends to be short ($\sim10^5$ years) enough to avoid the angular momentum loss before the core collapse.
}


The mass of the naked helium star is the sum of the primary helium core and the secondary helium core.
The radius of the naked helium star $R_{\rm rem}$ is calculated by Equation (81) in \cite{Hurley2000}.
{This formula is a fitting formula for the stellar evolution calculation of \cite{Pols1998}.}
 The synthesized naked helium star obtains a large angular momentum from the orbital angular momentum of the binary.
If the naked helium star rotates more rapid than the Kepler velocity, the outer part of the star is blown off. The orbital angular momentum is always larger than the spin angular momentum  of the Kepler velocity.
Thus, we assume that the naked helium star has a spin angular momentum of the Kepler velocity.
The resultant spin angular momentum is calculated by $kMR^2\Omega_{k}$, where $k$ and $\Omega_{k}$ are the momentum of inertia  and the angular Kepler spin velocity, respectively.  {We use the $k=0.21$, which is the the same value adopted in Hurley et al. (2000) for the dense convective core.}

{After the merger, the naked helium star loses the angular momentum by the stellar wind mass loss.}
For the wind mass-loss rate of the naked helium star, we adopt that for the Wolf-Rayet star,
\begin{equation}
\dot{M}_{\rm WR} =10^{-13}L^{1.5}\left(\frac{Z}{Z_{\odot}}\right) ^{0.86}~\rm M_{\odot}~yr^{-1}.
\end{equation}
This formula is a combination of the wind mass-loss rate in \cite{Hurley2000} and its metal dependence of Wolf-Rayet wind in \cite{Vink2005}.
 {To estimate the angular momentum loss of the naked helium star by the stellar wind mass loss \citep{Hurley2000}, we assume that the mass loss occurs uniformly at the stellar surface. In this case, the angular momentum loss of the stellar wind mass loss is calculated roughly as $\Delta{J}=-{2}/{3}(M_{\rm He,i}-M_{\rm BH})R^2\Omega_{\rm k}$, where $M_{\rm He,i}$ and $M_{\rm BH}$ are the initial mass of the merged naked helium star and the mass of BH, respectively.}

{Since the merger products are rapidly rotating, we assume that the naked helium stars evolve as the chemically homogeneous stars \citep{Maeder1987} and the entire naked helium star will  change into a CO star.}
When the stars collapse, we treat them as a direct collapse.
If the mass of the star after the collapse is larger than 3 $ M_{\odot}$, the star is regarded as a BH. 
A significant spin velocity just before the collapse may be required to induce a long GRB.
The condition for the normalized spin $a/M=cJ_{\rm preDC}/GM^2>1$ is simply taken as the criterion for the GRB progenitor in this paper.

We calculate the seven metallicity cases  as $Z=\zsun,~10^{-0.5}\zsun,~10^{-1}\zsun,~10^{-1.5}\zsun,~10^{-2}\zsun,~10^{-2.5}\zsun$ and $10^{-3}\zsun$.
For Pop I and II stars, our binary code can calculate a stellar evolution from $5\times10^{-3}\zsun$ to $1.5\zsun$, while the metallicity dependence of the wind mass loss is obtained for all the above ranges of $Z$.
The stellar evolution property below $5 \times 10^{-3} Z_\odot$ may be almost the same.
In order to calculate the cases for $Z=10^{-2.5}\zsun$ and $10^{-3}\zsun$, we combine the $5\times10^{-3}\zsun$ stellar evolution model and the wind mass-loss formulae for $Z<5 \times 10^{-3} Z_\odot$.

{Table \ref{IDF} shows the initial distribution functions.
We use the Salpeter mass function \citep[IMF$\propto M^{-2.35}$;][]{Salpeter1955}, and the other functions are determined from the Pop I massive binary observation \citep{Sana2012}.}
{In this calculation, we focus on the massive binaries, so we choose the minimum mass of the primary star as $5~\msun$.} 
Under this initial condition, we calculate $10^6$ binaries for each metallicity case.

In order to calculate the merger rate, we need the SFR and the metallicity evolution history.
We use the SFR {per comoving volume} in \cite{Madau2014}, hereafter MD14, as
\begin{equation}
SFR(z)=1.5\times10^{-2}\frac{(1+z)^{2.7}}{1+\left[\frac{1+z}{2.9}\right]^{5.6}}\rm M_{\odot}~yr^{-1}~Mpc^{-3},
\end{equation} 
above $0.1\msun$.
{The above SFR was obtained from the observation from $z=0$ to $z=8$.
As for the metallicity evolution, we use the mass-metallicity relation of galaxies \citep{Ma2016}
\begin{equation}
\log\left(\frac{Z}{Z_{\odot}}\right)=0.40\log\left(\frac{M_{\rm gal}}{10^{10}~\msun}\right)+0.67\exp(-0.5z)-1.04,
\end{equation}
and the Schechter function for the galaxy mass $M_{\rm gal}$ distribution,
\begin{equation}
\phi_{\rm sh}(M_{\rm gal})dM_{\rm gal}\propto\left(\frac{M_{\rm gal}}{M^*}\right)^\alpha\exp\left(-\frac{M_{\rm gal}}{M^*}\right)\frac{dM_{\rm gal}}{M^*}.
\end{equation} 
From $z=0$ to 4, the parameters in Equation (6) are taken to be
\begin{align}
\log M^*(z)=&11.16+0.17z-0.07z^2\\
\alpha(z)=&-1.18-0.082z.
\end{align}
They are  obtained by fitting the results in \cite{Fontana2004}.
From $z=4$ to 8, the parameters are assumed as
\begin{align}
\log M^*(z)=&10.72\\
\alpha(z)=&-1.508-0.176(z-4),
\end{align}
using data in \cite{Song2016}, {although there is  large uncertainty for such a high redshift.} 
We define the mass fraction of the galaxy whose metallicity is $Z$ as
\begin{equation}
f(Z,z)=\frac{\int^{{\rm max}[M_{\rm max}, M_{\rm gal}(10^{0.25}Z)]}_{{\rm min}[M_{\rm min}, M_{\rm gal}(10^{-0.25}Z)]}M_{\rm gal}\phi_{\rm sh}(M_{\rm gal})dM_{\rm gal}}{\int^{M_{\rm max}}_{M_{\rm min}}M_{\rm gal}\phi_{\rm sh}(M_{\rm gal})dM_{\rm gal}},
\end{equation}
where $M_{\rm max}=10^{12}~M_{\odot}$ and $M_{\rm min}=10^5~M_{\odot}$.
We evaluate the fractional SFR for each metallicity as
\begin{equation}
SFR_Z(Z,z)=f(Z,z)SFR(z).
\end{equation}
Finally,  the apparent long GRB rate for each metallicity is calculated as
\begin{equation}
R_{Z}(z)=f_{\rm B}\frac{f_{\rm b}}{1+f_{\rm b}}\frac{SFR_Z(Z,z)}{<M>}f_{\rm IMF}\frac{N_{Z,\rm GRB}}{N_{\rm total}},
\end{equation}
where $f_{\rm B}$ is the beaming factor of the GRB jets, $f_{\rm b}=0.5$ is the binary fraction, $<M>=\int^{100}_{0.1}M\cdot IMFdM=0.35~\msun$ is the average stellar mass, $f_{\rm IMF}=\frac{\int^{100}_5IMFdM}{\int^{100}_{0.1}IMFdM}=5\times10^{-3}$ is the normalization of IMF, and $N_{Z,\rm GRB}$ is the number of long GRBs in the metallicity $Z$ case obtained after $N_{\rm total}=10^6$ trial calculations of the binary evolution.
By summing up the rates for all metallicity cases, we obtain the final rate $R(z)$.
Note that the above assumption corresponds to the massive binary production rate of $4.7 \times 10^{-3} ({\rm SFR}/M_\odot ~\rm yr^{-1})~\mbox{yr}^{-1}$.

\begin{table*}[!t]
	\caption{The fractions of long GRBs, the fractions of case B, and the fractions of case C for each metallicity
	}
	\label{number}
	\begin{center}
		\begin{tabular}{ccccc}
			\hline
\multicolumn{1}{c}{}
      & \multicolumn{2}{c}{$\alpha\lambda=1$} & \multicolumn{2}{c}{$\alpha\lambda=0.1$}\\
			$Z$	& $N_{Z,\rm GRB}/N_{\rm total}$ [10$^{-3}$] &  [$cJ_{\rm preDC}/GM^2$] & $N_{Z,\rm GRB}/N_{\rm total}$ [10$^{-3}$]  & [$cJ_{\rm preDC}/GM^2$]\\
			\hline
              
			$\zsun$   &  0.385$\pm 0.020$ &40.4$\pm 10.4$ & 1.612$\pm 0.040$&42.8$\pm 8.3$ \\
			$10^{-0.5}\zsun$& 0.811$\pm 0.028$ &44.0$\pm 10.9$ & 3.465$\pm 0.059$ &45.5$\pm 7.9$ \\
			$10^{-1}\zsun$&  1.111$\pm0.033$ &46.2$\pm8.4$&5.162$\pm 0.072$&46.5$\pm6.3$ \\
			$10^{-1.5}\zsun$&  1.399$\pm 0.037$ &47.9$\pm 5.7$& 6.641$\pm 0.081 $& 46.8$\pm 5.5 $  \\
			$10^{-2}\zsun$& 2.137$\pm 0.046$&48.1$\pm 5.5$&7.893$\pm 0.089$  & 46.9$\pm5.6 $\\
			$10^{-2.5}\zsun$& 3.791$\pm 0.062$&49.0$\pm 4.8$&8.777$\pm 0.094$ & 47.4$\pm 5.7$ \\
			$10^{-3}\zsun$& 3.787$\pm 0.062$ &49.1$\pm 4.7$&8.771$\pm 0.094$  &47.5$\pm5.7 $\\
			\hline
		\end{tabular}
	\end{center}
\end{table*}
\section{Results}

{Table 2 shows the fractions of long GRBs and the average spins at pre-direct collapse for each metallicity and CE parameter.}
The fraction of long  GRBs increases as the metallicity decreases.
There are two reasons for this metallicity dependence.
First, when the wind mass loss is weak for a low metallicity system, 
{the binding energy loss of the binary is weak.}
Thus, the binaries tend to evolve as close binaries and they easily enter into a CE phase and merge.
Furthermore, the naked helium stars have difficulty losing the spin angular momentum.
The wind mass loss is almost negligible for $Z<10^{-2.5}\zsun$.

The second reason is the differences in the stellar evolution.
The low metallicity stars have such a large helium core mass that they more easily become BHs than the high metallicity stars \citep{Belczynski2010}.
Furthermore, the radius evolution depends on the metallicity.
{As shown in the examples in Figure 1, solar-metal primary stars may
lead to the CE phase during the HG phase or the RG phase
(the initial expansion phase), i.e. in the case B mass transfer. 
Unless the system is nearly equal mass binary, the secondary star at the CE phase may still be in the main-sequence (MS) stage. In this case, the merged star is regarded as a large MS star, which may lose its angular momentum via the stellar wind. As a result, the GRB production efficiency in case B becomes relatively low (Dewi et al. 2006).}

{On the other hand, if the CE phase is realized when the primary star is in the HeB or AGB phase, i.e. in case C mass transfer,
at which the secondary stars may already evolve to the post-MS stage.
In this case, the merged star may become a highly rotating helium star, which is preferable to cause a long GRB.
The fraction of the GRB progenitors in case B is always subdominant. Even in the high metallicity case ($Z>10^{-0.5} \zsun$), the fraction is 10-20 \% for $\alpha \lambda=1$.}

{For the low metallicity case, the occurrence rate of the CE in case C is higher, which enhances the GRB production efficiency (Podsiadlowski et al. 2010). The fraction of the case B GRBs is almost negligible for $Z<10^{-0.5}\zsun$.}
{As shown in Figure 1, low-metal stars tend to ignite the helium core burning (HeB) during the early evolutionary phase
because of their higher core temperature \citep{Pols1998}.
The resultant suppression of the radius in the HG phase makes it easier to avoid the undesirable CE in case B.}

{As shown in the Table 2, the number of long GRB progenitors in the $\alpha\lambda=0.1$ case is two to four times  the number  in the $\alpha\lambda=1$ case.
 {Low mass binaries are required to have a wide initial separation
to avoid the merge in the HG/RG phase, because their envelope
in the HG/RG phase is larger. Such low mass wide binaries can merge
in case C only for a lower $\alpha \lambda$.}
As a result, for a lower $\alpha \lambda$ even low mass binaries can merge in case C. The wind mass-loss rate of such low mass merged stars may be low enough even for their long lifetime. Therefore, a smaller $\alpha \lambda$ enhances the GRB rate especially in the high metallicity case.}


In terms of the property of the long GRB progenitors, the BH mass distribution almost traces the Salpeter IMF and is almost independent on the metallicity although the maximum mass is larger in  lower metallicity cases.
 {Therefore, the typical/average mass of the merged naked helium stars that launch a long GRB  is not as high as 3.5-5 $M_\odot$, for which the wind mass loss is not so efficient. This leads to the weak metallicity dependence of the final average spin as shown in Table 2.
The spin down due to the wind mass loss is effective only for the merged naked helium star whose mass is $>10 M_\odot$ for the solar metallicity. However, the fraction of such high mass stars is relatively small.}

\begin{figure}[!h]
		\includegraphics[width=0.5\textwidth,clip=true]{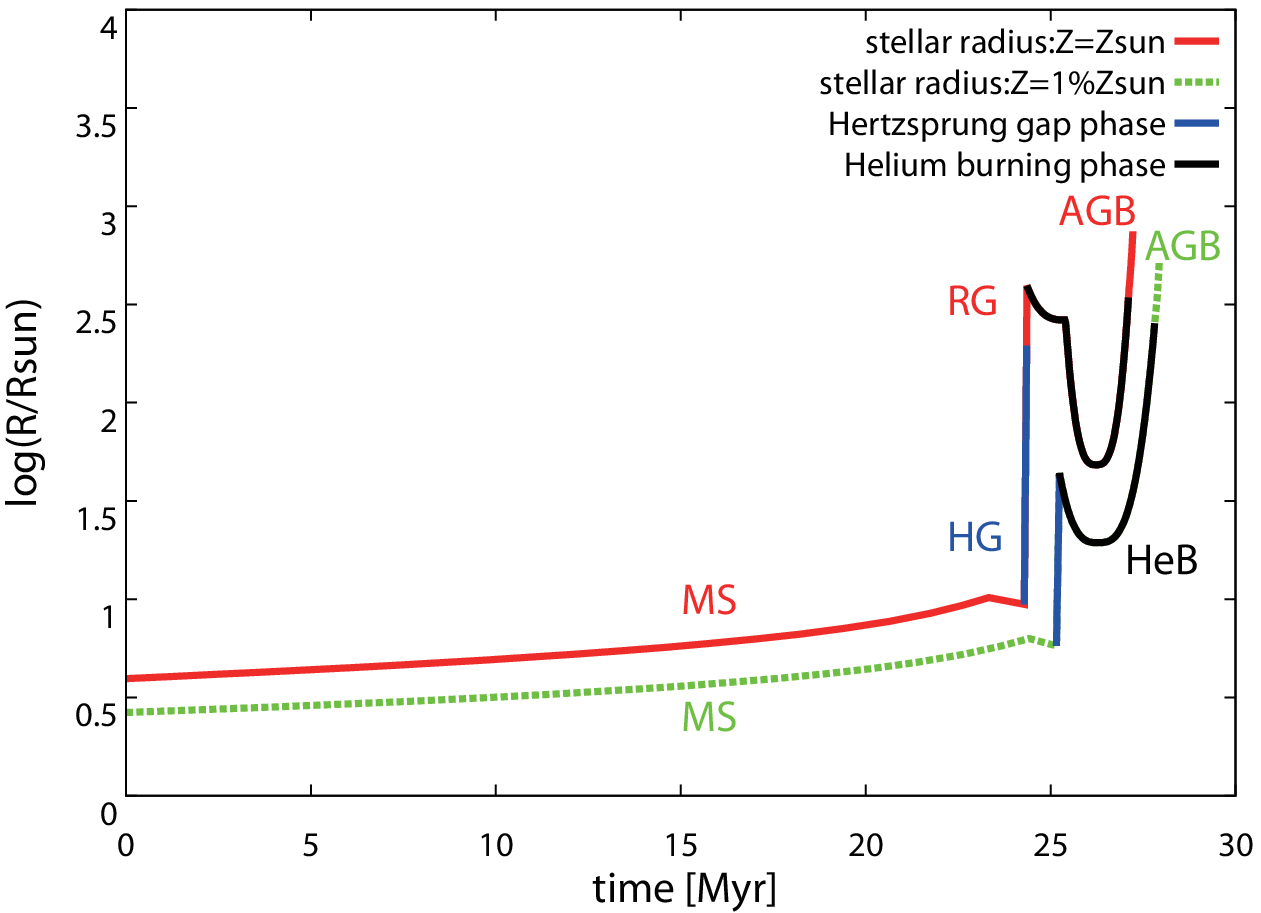}
	\caption{Radius evolutions of $10~\msun$ stars for $ Z=\zsun$ and $Z=0.01\zsun$.}
	\label{fig:GRB}
\end{figure}

Figure 2 shows the long GRB rate in our results (black lines).
The absolute value of the GRB rate in our estimate depends on the uncertain beaming factor. Within a reasonable range of $f_{\rm B}$, the rate is consistent with the results in WP10 and L14.
 {Our results keep a higher rate even at high redshifts compared to the SFR, which is due to the higher fraction of GRBs at lower metallicity.
This tendency seems qualitatively consistent with the GRB rates in WP10 and L14.
The long GRB rate in WP10 was directly obtained for each redshift bin.
On the other hand, L14 postulated a functional form of the redshift evolution of the GRB rate, and obtained the parameters in the function. Therefore, while the best-fit function was obtained as shown in Fig. 2, the error for each redshift was not directly provided.
Our results may be in the uncertainty of the GRB rate history.

Our model has difficulty  agreeing with the peak redshift $z=3.6$ in L14. A more prominent metallicity evolution is required to reproduce the peak in L14.
Alternatively, the spin evolution shown in Table 2 may affect the LF or beaming factor. Even if the evolutions of the LF and/or $f_{\rm B}$ are not so stronger than the assumptions in some previous studies, the peak redshift in the apparent GRB rate in L14 may be biased by those evolutions.
}

\begin{figure}
		\includegraphics[width=0.5\textwidth,clip=true]{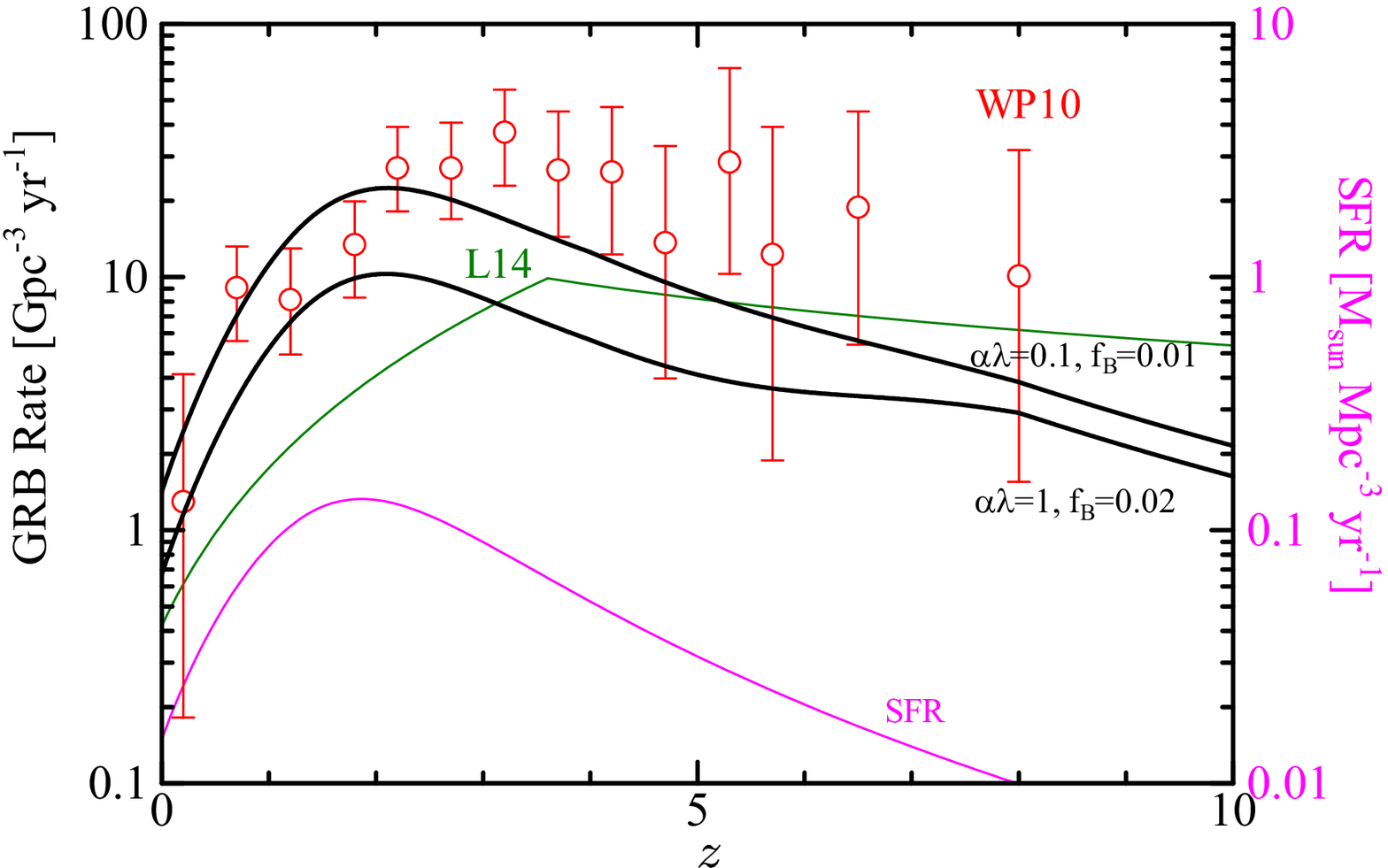}
	\caption{Apparent long GRB rates in our model (black lines). The upper solid line is the result for $\alpha\lambda=0.1$ and the beaming factor $f_{\rm B}=0.01$, while the lower solid lines show the case of $\alpha\lambda=1$ and $f_{\rm B}=0.02$. The estimated rates from the {\it Swift} samples are also shown by the red data points (WP10) and the  green line (L14). The purple line (see the right axis) shows the SFR in MD14.}
	\label{fig:GRB}
\end{figure}

\section{Conclusion and Discussion}

We consider the binary merger progenitor model to be the long GRB progenitor.
At high redshifts, the fraction of binary merger progenitors is larger than that of low redshift  due to the metallicity evolutions of the wind mass loss and stellar evolution.
The GRB rate in this scenario is roughly consistent with the observed rate,
in spite of the some simple assumptions: constant $f_{\rm b}$ and $f_{\rm B}$, fixed initial distribution functions, and simplified mass-metallicity relation of galaxies, and so on.
In addition, our results reproduce the relatively higher GRB rate at high redshifts compared to the SFR without the evolutions of the LF or beaming factor, though both the model and observations still have large uncertainty. The metallicity evolution may affect not only the GRB rate but also the GRB property, which will be a future theme to further reconcile the observations and models.

Future X-ray missions, such as HiZ-GUNDAM (Yoshida et al. 2016)
and THESEUS (Yuan et al. 2016) may detect GRBs for $z>8$.
Here we consider a mission with the detection threshold of the flux
$10^{-9}~\mbox{erg}~\mbox{cm}^{-2}~\mbox{s}^{-1}$
at 0.3--5 keV, field of view of 1.8 sr, and a 100 \% duty cycle.
To estimate the flux in the X-ray band, the GRB spectrum
is assumed to be the Band function with $\alpha=-1$,
$\beta=-2.25$, and the modified Yonetoku relation for the spectral peak energy
adopted in L14.
Although the SFR and metallicity evolution for $z>8$ are highly unknown,
here we simply extrapolate the equation (2) for the SFR
and assume $f(Z,z)=f(Z,8)$ for $z>8$.
Adopting the best-fit model in L14 for the GRB LF,
which does not evolve with redshift,
the GRB detection rate with the above instrument is expected to
be  {24.7 (8.7) events per year for $z>8$ (10) for $\alpha\lambda=0.1$ and $f_{\rm B}=0.01$.}
The LF in WP10 includes a larger fraction
of dim GRBs than that in L14, so that the detection rate
estimated with the LF in WP10 is suppressed by a factor of $\sim 2$.

Although the functional shape of the GRB rate $R(z)$ is mainly determined by the metallicity evolution, its absolute value can be magnified by taking a small $\alpha \lambda$ parameter.
Alternatively, the chemically homogeneous evolution by the tidal effect increases the fraction of long GRB progenitors \citep{Cantiello2007}.
If a binary is initially so close, the spin velocity of the stars nearly become the Keplar velocity at the MS phase by the tidal effect.
In this case, the stars have high spins and can perform the chemically homogeneous evolution.
For simplicity, we have not included this possible effect at the MS stage in our calculation.
Low metallicity binaries are easier to satisfy such a condition than the solar-like metallicity binaries thanks to their weak stellar wind mass loss.

{Even if the AGB star does not fulfill the Roche lobe, the matter of slow winds possibly fulfills the Roche lobe, and mass transfer probably occurs for a larger orbital period than the simplest stellar models predict  \cite[sometimes referred to as the case D
mass transfer][]{P2007}.
If the mass transfer in the case D leads to the CE phase, the number of long GRB progenitors is possibly enhanced by this effect..}

\section*{Acknowledgment}
We thank the anonymous referee, Takatoshi Shibuya, and Akira Konno for discussions and useful comments.
~~This work was supported by JSPS KAKENHI Grant Number
JP16818962 (TK) and Grants-in-Aid for Scientific Research Nos. 15K05069 and
16K05291 (KA).


\end{document}